\documentclass{article}
\usepackage{amssymb,amsmath}
\usepackage{epsfig}
\usepackage{pdflscape}
\usepackage{subfigure}
\usepackage{color}
\usepackage{lscape}
\usepackage{bbm}
\usepackage{bm}
\usepackage{mathrsfs}
\usepackage{amsfonts}
\usepackage{graphics}
\usepackage{indentfirst}
\usepackage[square,numbers,sort&compress]{natbib}
\usepackage{multicol}

\newtheorem{rem}{Remark}[section]
\newcommand{\br}{\begin{rem}}
\newcommand{\er}{\end{rem}}
\newtheorem{ex}[rem]{Example}
\newcommand{\bex}{\begin{ex}}
\newcommand{\eex}{\end{ex}}
\newtheorem{Def}[rem]{Definition}
\newcommand{\bd}{\begin{Def}}
\newcommand{\ed}{\end{Def}}
\newtheorem{theorem}[rem]{Theorem}
\newcommand{\bt}{\begin{theorem}}
\newcommand{\et}{\end{theorem}}
\newtheorem{Prop}[rem]{Proposition}
\newcommand{\bp}{\begin{Prop}}
\newcommand{\ep}{\end{Prop}}
\newtheorem{lemma}[rem]{Lemma}
\newcommand{\bl}{\begin{lemma}}
\newcommand{\el}{\end{lemma}}
\newcommand{\be}{\begin{equation}}
\newcommand{\ee}{\end{equation}}
\newcommand{\bea}{\begin{eqnarray}}
\newcommand{\eea}{\end{eqnarray}}
\newcommand{\pa}{\partial}
\newcommand{\nn}{\nonumber}
\newcommand{\adots}{\mathinner{\mkern2mu\raise1pt\hbox{.}\mkern2mu
\raise4pt\hbox{.}\mkern2mu\raise7pt\hbox{.}\mkern1mu}}

\title{Scaling Symmetry Reductions of\\ Coupled KdV Systems}

\author{Allan P. Fordy, School of Mathematics,\\
University of Leeds, Leeds LS2 9JT, UK.\\ ~~E-mail: a.p.fordy@leeds.ac.uk}

\begin{document}

\maketitle

\begin{abstract}
In this paper we discuss the Painlev\'e reductions of coupled KdV systems.  We start by comparing the procedure with that of {\em stationary reductions}.  Indeed, we see that exactly the same construction can be used at each step and parallel results obtained.  For simplicity, we restrict attention to the $t_2$ flow of the KdV and DWW hierarchies and derive respectively 2 and 3 compatible Poisson brackets, which have identical {\em structure} to those of their stationary counterparts.  In the KdV case, we derive a discrete version, which is a non-autonomous generalisation of the well known Darboux transformation of the stationary case.
\end{abstract}



\section{Introduction}

In 1976 Bogoyavlenskii and Novikov  \cite{76-5} considered stationary flows of the KdV hierarchy and showed that each stationary flow defines a completely integrable, finite dimensional Hamiltonian system. Importantly the Lagrangian and canonical coordinates of the stationary flow, as well as the first integrals, were {\em deduced} from those of the KdV hierarchy.
This idea was generalised to {\em coupled KdV (cKdV)} hierarchies, as well as to higher order Lax operators, in several papers (see, for example, \cite{f87-3,f91-1,f95-2,f95-3,00-4}).  After some years of dormancy, the subject has recently been revisited \cite{f23-1,23-2,24-1,f24-1}.  The emphasis in \cite{23-2,24-1} has been on separability, whilst that of \cite{f23-1,f24-1} has been to show how the important structures of the stationary systems are {\em directly constructed} from those of the original PDE system.

In this paper we consider (some specific) nonlinear evolution equations of the cKdV hierarchy.  Recall that these are isospectral to the ``energy-dependent Schr\"odinger operator'' \cite{f87-5,f89-2}
\be\label{Scrod-op}
L  = \pa_x^2 + u, \quad\mbox{with}\quad u = \sum_{i=0}^N u_i \lambda^i, \;\;\mbox{where}\;\;  u_N = -1,
\ee
and that the resulting $N$ component system for $(u_0,\dots ,  u_{N-1})$ has $N+1$ compatible Hamiltonian operators, with each flow being written
\be\label{Ncpt-HamRep}
{\bf u}_{t_n} = B_N \delta_u H_n = B_{N-1} \delta_u H_{n+1} = \cdots = B_0 \delta_u H_{n+N}.
\ee
In \cite{f89-2} it was shown that the $N$ component system possessed $N$ modifications, generalising the MKdV hierarchy.  These play an important role in this paper.

These equations also have some additional Lie point symmetries: time translation and scaling.  For each of these we can consider invariant solutions, which are therefore constrained to satisfy an additional first order, linear PDE, with the result that (\ref{Ncpt-HamRep}) reduces to an ODE.  It is not always the case that a Hamiltonian formulation can be reduced to this ODE.

The simplest case is the {\em stationary flow}, ${\bf u}_t=0$, related to the time-translation symmetry.  If
\be\label{ut=BdH}
{\bf u}_t = B \delta_u H,
\ee
and $\mbox{Ker} B=\delta_u {\cal C}$, for some Casimir $\cal C$, then
\be\label{BdH=0}
B \delta_u H = 0 \quad\Rightarrow\quad   \delta_u \left(H-{\cal C}\right) =0,
\ee
giving a Lagrangian form of the reduced equation for the stationary flow, from which a canonical Hamiltonian formulation can be found.  This is the case discussed in the papers listed above.

In this paper we consider the reduction related to the scaling symmetry, which gives rise to equations of ``Painlev\'e'' type (see \cite{77-2,98-4,01-7}).  For this reduction, most of the Hamiltonian formulations cannot be reduced.  For example, the KdV operator $B_0$ and the dispersive water wave (DWW) operators $B_0$ and $B_2$ cannot be reduced.  However, we still have a Miura map to a modified system, so, when (\ref{ut=BdH}) can be reduced, we can still construct multi-Hamiltonian formulations for the finite dimensional, reduced system.  In fact the formulae can be taken directly from the stationary case.

It was shown in \cite{f23-1,f24-1} that the second nontrivial flow (the ``$t_2$ flow'') has (for each $N$) a particularly interesting stationary reduction, of the form
\be\label{Gen-t2-reduc}
h^{(Q)} = \frac{1}{2} (P_1^2+P_2^2) + U(Q_1,Q_2),
\ee
where $U(Q_1,Q_2)$ is one of the family of potentials, separable in {\em parabolic coordinates} (see Equation 2.2.41 in \cite{90-16}):
\be\label{parabolic-uv}
u = \sqrt{Q_1^2+Q_2^2}+Q_1,\quad v = \sqrt{Q_1^2+Q_2^2}-Q_1.
\ee
We consider non-autonomous extensions of these, arising from the scaling symmetry reductions.  Some of these equations are well known (see \cite{98-4,01-7}), but here we also introduce generalisations of the Poisson brackets derived in \cite{f23-1,f24-1}.  In fact, it was shown in \cite{f24-1} that the entire class of $h^{(Q)}$ (separable in parabolic coordinates) have Poisson brackets which are written in terms of gradients of $h^{(Q)}$ and the commuting integral $f^{(Q)}$.  These formulae do not change in the non-autonomous case, but the functions $h^{(Q)},\, f^{(Q)}$ now possess non-autonomous terms.

This brief paper presents these ideas in two main sections.

Section \ref{sec:KdV} is concerned with the simplest case of (\ref{Scrod-op}) with $N=1$, which is the standard KdV hierarchy.  For the $t_2$ flow, this gives a non-autonomous H\'enon-Heiles system (see also \cite{98-4}) and its modification.  The Miura map is used to construct a bi-Hamiltonian formulation of both.
In \cite{f95-1} the Darboux transformation is used to construct a discrete version of the stationary flows of the MKdV hierarchy.  This is generalised to the non-autonomous case in Section \ref{sec:discrete}.

Section \ref{sec:DWW} is concerned with the case of (\ref{Scrod-op}) with $N=2$, which is the DWW hierarchy.  For the $t_2$ flow, this gives a non-autonomous Hamiltonian with quartic potential (see also \cite{01-7}).  The Miura map is now used to construct a tri-Hamiltonian formulation of this and its modification.

\section{The KdV and MKdV Hierarchies (N=1)}\label{sec:KdV}

This is a well known case, with $u=u_0-\lambda$, so some details will be sparse.

\subsection{The KdV Hierarchy}

The KdV hierarchy has the bi-Hamiltonian representation
\begin{subequations}
\be\label{KdV-utn}
u_{0t_n} =  B_1^u \delta_{u_0} H_n^u = B_0^u \delta_{u_0} H_{n+1}^u,\;\; n\geq 0,
\ee
with
\be\label{KdV-B1B0}
B_1^u = \pa_x^3+4 u_0 \pa_x+2 u_{0x},\quad B_0^u = 4 \pa_x ,
\ee
where $\delta_{u_0}$ denotes the variational derivative with respect to $u_0$.  The first 4 Hamiltonian densities are
\be\label{KdV-Hm}
 H_0^u =   u_0,\;\;\; H_1^u = \frac{1}{4}\, u_0^2,\;\;\; H_2^u = \frac{1}{8} \left(u_0^3-\frac{1}{2} u_{0x}^2\right),\;\;\; H_3^u = \frac{1}{64} \left(5 u_0^4-10 u_0 u_{0x}^2+u_{0xx}^2\right).
\ee
\end{subequations}

\subsubsection{Reduction Through Scaling Symmetry}

The scaling symmetry for the $t_n$ flow is
\begin{subequations}
\be\label{KdV-S}
S^u = x \pa_x +(2n+1) t_n \pa_{t_n}-2 u_0 \pa_{u_0}, \;\;\;\mbox{with invariants}\;\;\; z= x\, t_n^{-\frac{1}{2n+1}},\;\; w = t_n^\frac{2}{2n+1} u_0.
\ee
We substitute $u_0(x,t_n) = w(z)\, t_n^{-\frac{2}{2n+1}}$ and find
\be\label{dzB1w}
\pa_x = t_n^{-\frac{1}{2n+1}}\, \pa_z,\;\; B_1^u = t_n^{-\frac{3}{2n+1}} B_1^w, \;\; H_\ell^u = t_n^{-\frac{2(\ell+1)}{2n+1}} H_\ell^w,\;\; \delta_{u_0} = t_n^{\frac{2}{2n+1}}\, \delta_w,
\ee
where, for $B_1^w$ and $H_n^w$, we just replace $u_0$ by $w$ (and $\pa_x$ by $\pa_z$).  Since this is a symmetry of the flow, the result of substitution in (\ref{KdV-utn}) has a {\em homogeneous} factor of $t_n^{-\frac{2n+3}{2n+1}}$, which can be cancelled, so the result can be written
\be\label{KdV-wtn}
-\frac{1}{2n+1} \, \left(z w_z+2 w\right) = B_1^w \delta_w H_n^w = B_0^w \delta_w H_{n+1}^w.
\ee
Since $z w_z+2 w$ is not in the range of $B_0^w = 4\pa_z$, we cannot write a formula analogous to (\ref{BdH=0}).  However
\be\label{B1z}
B_1^w z = 4w+2zw_z \quad\Rightarrow\quad  B_1 \delta_w \left(H_n^w+\frac{z w}{2 (2n+1)} \right) = 0.
\ee
Following the approach outlined in \cite{f95-3} and \cite{f23-1} (after Eq (3.2a)), we have
\be\label{kdv-ham1w-phi}
\delta_w \left( H_n^w + \frac{z w}{2(2n+1)}\right) = \frac{1}{2} \varphi^2 \quad\Rightarrow\quad  \varphi_{zz} + w\varphi = 2 \beta \varphi^{-3},
\ee
leading to the Lagrangian
\be\label{kdv-phi-lag}
{\cal L}_n^w = H_n^w + \frac{z w}{2(2n+1)} -\frac{1}{2} w \varphi^2 + \frac{1}{2} \varphi_z^2-\frac{\beta}{\varphi^2} .
\ee
\end{subequations}

\subsection{The Miura Map and MKdV Hierarchy}

The modified KdV hierarchy is related to the KdV hierarchy through the Miura map
\begin{subequations}
\be\label{kdv-miura}
u_0 = -v_{0x}-v_0^2,
\ee
and the second Hamiltonian operator, $B_1$, of the KdV hierarchy factorises
\be\label{mkdv-v0tn}
\pa_x^3+4 u_0 \pa_x+2 u_{0x} = (-\pa_x-2 v_0) (-\pa_x) (\pa_x-2 v_0) \quad\Rightarrow\quad  v_{0t_n} = -\pa_x \, \delta_{v_0} H_n^v,
\ee
\end{subequations}
where $H_n^v[v_0] = H_n^u\left[-v_{0x}-v_0^2\right]$.  There is an additional Hamiltonian density $H_{-1}^v = v_0$, which is the Casimir of the Hamiltonian operator.  Also, by direct substitution, the formula $H_n^u\left[-v_{0x}-v_0^2\right]$ includes {\em exact $x-$derivatives}, which can be removed, to simplify the formula for $H_n^v[v_0]$.

\subsubsection{Reduction through Scaling Symmetry}

Here we use $z= x\, t_n^{-\frac{1}{2n+1}}$ and $v_0(x,t_n) = \theta(z)\, t_n^{-\frac{1}{2n+1}}$ (which follows from the Miura map), leading to
\begin{subequations}
\be\label{mkdv-ham1phi}
v_{0t_n} = \frac{-1}{(2n+1) t_n^{\frac{2(n+1)}{2n+1}}} \left(z \theta\right)_z,\quad H_n^v = t_n^{-\frac{2(n+1)}{2n+1}} H_n^\theta,\quad
                     \delta_{v_0} = t_n^{\frac{1}{2n+1}}\, \delta_\theta,\quad \pa_x = t_n^{-\frac{1}{2n+1}} \pa_z.
\ee
Piecing these together, we obtain
\be\label{mkdv-hamtheta}
 \frac{-1}{(2n+1)}\, \pa_z\left(z \theta\right) = -\pa_z \delta_\theta H_n^\theta   \quad\Rightarrow\quad
         \delta_\theta {\cal L}_n^\theta = 0, \;\;\mbox{where}\;\; {\cal L}_n^\theta  = H_n^\theta - \frac{1}{2(2n+1)}z \theta^2-\alpha \theta.
\ee
\end{subequations}

\subsection{The case $n=2$: Non-autonomous H\'enon-Heiles system and Modification}

Here we consider this specific case of the Lagrangians (\ref{kdv-phi-lag}) and (\ref{mkdv-hamtheta}) and the related Hamiltonian systems, together with the Poisson map between them.  This corresponds to the H\'enon-Heiles system $H_{(ii)}$ in Section 3.2 of \cite{98-4}.

We also consider a {\em discrete version} of the MKdV case.

\subsubsection{The Non-autonomous H\'enon-Heiles system}

Using $-8 H_2^w$ in (\ref{kdv-phi-lag}), we obtain
\begin{subequations}
\be\label{kdv-phi-lag-n=2}
{\cal L}_2^w = \frac{1}{2} (w_z^2+\varphi_z^2) -\frac{1}{2} w (2 w^2+\varphi^2) -\frac{\beta}{\varphi^2} + \frac{z w}{10}.
\ee
With $w=Q_1,\; \varphi = Q_2,\; w_{z}=P_1,\; \varphi_z = P_2$, we obtain the Hamiltonian
\be\label{kdv-n=2-hQ}
h^{(Q)} = \frac{1}{2}(P_1^2+P_2^2)+\frac{1}{2} Q_1 \left(2 Q_1^2+Q_2^2\right) +\frac{\beta}{Q_2^2} - \frac{1}{10}z Q_1 ,
\ee
which is a non-autonomous version of (3.13) in \cite{f23-1} (see also \cite{98-4}).  We can then generalise the first integral (3.15c) (of \cite{f23-1}) to obtain
\be\label{kdv-n=2-fQ}
f^{(Q)} = P_2 (Q_2 P_1-Q_1 P_2) +\frac{1}{8} Q_2^2 (4 Q_1^2+Q_2^2) - \frac{2 \beta Q_1}{Q_2^2} - \frac{1}{20}\, z Q_2^2
\ee
which Poisson commutes with $h^{(Q)}$.
\end{subequations}

\subsubsection{The Non-autonomous Modified H\'enon-Heiles system}

In (\ref{kdv-phi-lag}), we used $H_2^w = \frac{1}{2} w_z^2-w^3$, to get Lagrangian (\ref{kdv-phi-lag-n=2}), giving (up to exact derivative) $H_2^\theta = \frac{1}{2} \theta_{zz}^2+5 \theta^2\theta_z^2+\theta^6$, so using (\ref{mkdv-hamtheta}) as a Lagrangian, we find
\bea
&&  q_1=\theta,\;\; q_2=\theta_z,\;\; p_2= \theta_{zz}, \;\; p_1 = \frac{\pa {\cal L}_2^\theta}{\pa \theta_z} -p_{2z},\nn\\[-2mm]
&&         \label{mkdv-n=2-lag+ham}    \\[-2mm]
&&   h^{(q)} = q_{1z} p_1+q_{2z} p_2-{\cal L} = \frac{1}{2}p_2^2+q_2 p_1-5 q_1^2 q_2^2-q_1^6+\frac{1}{10} z q_1^2+\alpha q_1.   \nn
\eea

\subsubsection{The Poisson Map}

We start with the pair of Poisson commuting Hamiltonians (\ref{kdv-n=2-hQ}) and (\ref{kdv-n=2-fQ}).  The Poisson map is built from the Miura map (\ref{kdv-miura}), using the Hamiltonian $h^{(Q)}$:
\begin{subequations}\label{Pmap-gen-n=2}
\bea
&&  Q_1 = -q_2-q_1^2,  \label{Pmap-Q1-n=2}\\
&&  Q_{1zz} = \{\{Q_1,h^{(Q)}\},h^{(Q)}\} \;\;\;\Rightarrow\;\;\;      Q_2 = \sqrt{\frac{z}{5} -6 Q_1^2-2 Q_{1zz}},  \label{Pmap-Q2-n=2}\\
&&  P_1=Q_{1z},\quad P_2 = Q_{2z},  \label{Pmap-Pi-n=2}\\
&& Q_{2zz} = \{\{Q_2,h^{(Q)}\},h^{(Q)}\} \;\;\;\Rightarrow\;\;\;   \beta = \frac{1}{2} Q_2^3 (Q_1 Q_2 + Q_{2zz}).   \label{Pmap-b-n=2}
\eea
\end{subequations}
To calculate the $z-$derivatives in terms of $\bf q$ (starting with (\ref{Pmap-Q1-n=2}) and (\ref{Pmap-Q2-n=2})), we use (for any function $g(z)$) $g_z = \{g({\bf q}),h^{(q)}\}$, where $h^{(q)}$ is given by (\ref{mkdv-n=2-lag+ham}).

\medskip
Defining $A^\pm = p_1 \pm 2 q_1 p_2 \mp 3 q_1^4-4 q_1^2 q_2 \mp q_2^2 \pm \frac{z}{10}$, the map (\ref{Pmap-gen-n=2}) takes the specific form
\begin{subequations}
\be \label{Pmap-n=2}
Q_1 = -q_2-q_1^2,\;\; Q_2 = \sqrt{-2 A^-},\;\; P_1 = -p_1 -2 q_1 q_2, \;\; P_2 = \frac{\alpha}{\sqrt{-2 A^-}} + q_1 \sqrt{-2 A^-},\;\; \beta = -\frac{1}{2} \alpha^2,
\ee
leading to
\be\label{Pmap-fq-n=2}
h^{(Q)} \mapsto h^{(q)},\quad      f^{(Q)} \mapsto f^{(q)} = \frac{1}{2} A^+ A^- -\alpha (p_2-2 q_1^3).
\ee
\end{subequations}

\medskip
The Hamiltonian $h^{(Q)}$ of (\ref{kdv-n=2-hQ}) is an example of (57d) in \cite{f24-1} (with $\frac{z}{10}$ being the parameter $\kappa_2$), so we can write the Poisson brackets:
\begin{subequations}
\be\label{PBQ01-n=2}
{\cal{P}}_0^{(Q)} = \left(\begin{array}{ccccc}
  0&0&0&\frac1{Q_2} &a_{15}    \\[2mm]
  0&0&\frac1{Q_2}&-\frac{2Q_1}{Q_2^2} & a_{25}   \\[2mm]
  0&-\frac1{Q_2}&0&\frac{P_2}{Q_2^2} & a_{35}   \\[2mm]
  -\frac1{Q_2}&\frac{2Q_1}{Q_2^2}&-\frac{P_2}{Q_2^2}&0  & a_{45}  \\[2mm]
   -a_{15} & -a_{25} &-a_{35} & -a_{45} &0
\end{array}\right) , \qquad
{\cal{P}}_1^{(Q)} =
\left(\begin{array}{rrrrr}
0&0&1&0&0  \\
0&0&0&1&0  \\
-1&0&0&0&0 \\
0&-1&0&0&0 \\
0&0&0&0&0
\end{array}\right),
\ee
where the column $(a_{15},a_{25},a_{35},a_{45},0)^T=-P_1^{(Q)}\nabla_Q f^{(Q)}$.  The 3 functions $\beta,\; h^{(Q)}$ and $f^{(Q)}$ satisfy:
\bea
&&  {\cal{P}}_1^{(Q)} \nabla_Q \beta = {\cal{P}}_0^{(Q)} \nabla_Q h^{(Q)} = 0,  \nn\\[-1mm]
&&     \label{biHamQ-n=2}  \\[-1mm]
&&  {\bf Q}_{t_h} = {\cal{P}}_0^{(Q)} \nabla_Q f^{(Q)} = {\cal{P}}_1^{(Q)} \nabla_Q h^{(Q)},\quad {\bf Q}_{t_f} = {\cal{P}}_0^{(Q)} \nabla_Q (-\beta) = {\cal{P}}_1^{(Q)} \nabla_Q f^{(Q)}.\nn
\eea
These are identical to formulae given on p19 of \cite{f23-1}, but now they contain non-autonomous terms, hidden away in the definitions of $h^{(Q)}$ and $f^{(Q)}$.

We then use the Poisson map (\ref{Pmap-n=2}) to construct ${\cal{P}}_i^{(q)}$, with ${\cal{P}}_i^{(Q)}=\left(\frac{\pa {\bf Q}}{\pa {\bf q}}\right) {\cal{P}}_i^{(q)} \left(\frac{\pa {\bf Q}}{\pa {\bf q}}\right)^T$, where ${\cal{P}}_1^{(q)}={\cal{P}}_1^{(Q)}$ (canonical transformation) and
\be\label{PBq0-n=2}
{\cal{P}}_0^{(q)}= \frac{1}{\alpha} \left(\begin{array}{ccccc}
                               0 & -q_1 & p_2-6 q_1^3 & -q_2 & c_{15}  \\
                               q_1 & 0 & 8 q_1^2 q_2 & 4 q_1^3 & c_{25} \\
                               6 q_1^3-p_2 & -8 q_1^2 q_2 & 0 & c_{34} & c_{35} \\
                               q_2 & -4 q_1^3 & -c_{34} & 0 & c_{45} \\
                               -c_{15} & -c_{25} & -c_{35} & -c_{45} & 0
                               \end{array}\right)
\ee
where $c_{34} = \alpha-4 q_1^2 p_2+24 q_1^5-8 q_1 q_2^2$ and $(c_{15},c_{25},c_{35},c_{45},0)^T=P_1^{(q)}\nabla_q f^{(q)}$.  The 3 functions $\frac{1}{2} \alpha^2,\; h^{(q)}$ and $f^{(q)}$ satisfy:
\bea
&&  {\cal{P}}_1^{(q)} \nabla_q \left(\frac{1}{2} \alpha^2\right) =  {\cal{P}}_0^{(q)} \nabla_q h^{(q)} = 0,  \nn\\[-1mm]
&&     \label{biHamq-n=2}  \\[-1mm]
&&  {\bf q}_{t_h} = {\cal{P}}_0^{(q)} \nabla_q f^{(q)} = {\cal{P}}_1^{(q)} \nabla_q h^{(q)},\quad {\bf q}_{t_f} = {\cal{P}}_0^{(q)} \nabla_q \left(\frac{1}{2} \alpha^2\right) = {\cal{P}}_1^{(q)} \nabla_q f^{(q)}.\nn
\eea
\end{subequations}

\subsection{The Discrete Case}\label{sec:discrete}

In \cite{f95-1} (see Section 4.2) we used a Darboux transformation to construct a discrete version of the stationary flows of the MKdV hierarchy.
It is straightforward to generalise these maps to the non-autonomous case.  Recall that we have two choices of Miura map (\ref{kdv-miura}), giving $Q_1=-\tilde q_2-\tilde q_1^2=q_2-q_1^2$, where $q_1$ and $\tilde q_1$ correspond to two different solutions of (\ref{mkdv-n=2-lag+ham}).  We use this and its differential consequences, $\tilde p_2 +p_2+2 q_1 q_2-2 \tilde q_1 \tilde q_2 = 0$, etc, to construct the equations, which can be solved explicitly, to give (with $\alpha$ fixed)
\bea
&&  \tilde q_1 = -q_1 -\frac{\alpha}{A^+}, \qquad \tilde q_2 = -q_2-\frac{2 \alpha q_1}{A^+} - \frac{\alpha^2}{(A^+)^2},  \nn  \\
&&  \tilde p_1 = -p_1 +\frac{2\alpha (p_2-6 q_1^3-4 q_1 q_2)}{A^+} - \frac{2\alpha^2 (11 q_1^2+q_2)}{(A^+)^2} - \frac{16 \alpha^3 q_1}{(A^+)^3} - \frac{4 \alpha^4}{(A^+)^4},  \label{qtilde-ptilde}\\
&&  \tilde p_2 = - p_2 - \frac{2 \alpha (2 q_1^2+q_2)}{A^+} - \frac{6 \alpha^2 q_1}{(A^+)^2}  - \frac{2 \alpha^3}{(A^+)^3} ,  \nn
\eea
where $A^+$ is defined above (\ref{Pmap-n=2}).  This map preserves both Poisson brackets and the functions $h^{(q)}$ and $f^{(q)}$ and satisfies $A^+({\bf q})=-A^-({\bf \tilde q})$.

\br
This construction relies on the discrete symmetry $v\mapsto -v$ of the MKdV hierarchy.
\er

\section{The DWW and MDWW Hierarchies (N=2)}\label{sec:DWW}

This is the simplest example of a (truly) {\em coupled} KdV system.

\subsection{The DWW Hierarchy}

The DWW hierarchy is tri-Hamiltonian, with the $n^{th}$ flow being written
\begin{subequations}
\be\label{dww-triHam}
 {\bf u}_{t_n} = B_2 \delta_u H_n = B_1 \delta_u H_{n+1} = B_0 \delta_u H_{n+2} ,
\ee
where
\be \label{dww-Bi}
                 B_2 = \left(
                                \begin{array}{cc}
                                 0 & J_0^u \\
                                 J_0^u & J_1^u
                                   \end{array}
                                 \right),\;\; B_1 = \left(
                                                        \begin{array}{cc}
                                                         J_0^u & 0 \\
                                                          0 & -J_2^u
                                                          \end{array}
                                                          \right),\;\; B_0 = \left(
                                                                                 \begin{array}{cc}
                                                                                  -J_1^u & -J_2^u \\
                                                                                   -J_2^u & 0
                                                                                   \end{array}
                                                                                   \right),
\ee
with ${\bf u} = (u_0,u_1)^T$ and
\be\label{dww-Jiu}
J_0^u = \pa_x^3+4 u_0 \pa_x+2 u_{0x} ,\quad J_1^u = 4 u_1 \pa_x+2 u_{1x}, \quad J_2^u = -4 \pa_x.
\ee
The first 4 Hamiltonians are
\be\label{dww-Hn}
H_0^u = u_1,\;\;\; H_1^u = u_0+\frac{1}{4} u_1^2,\;\;\; H_2^u = \frac{1}{2} u_1\left(u_0+\frac{1}{4} u_1^2\right),\;\;\; H_3^u = \frac{1}{4}\left(u_0^2+\frac{3}{2} u_0u_1^2 +\frac{5}{16} u_1^4-\frac{1}{4}u_{1x}^2\right).
\ee
\end{subequations}

\subsubsection{Reduction Through Scaling Symmetry}

The DWW hierarchy corresponds to the Lax operator (\ref{Scrod-op}) with $N=2$, with scaling symmetry
\begin{subequations}
\be\label{dwwscale}
S^u = x \pa_x+ m t_n \pa_{t_n} -2 u_0 \pa_{u_0} -u_1 \pa_{u_1},\quad\mbox{and invariants}\;\; z = \frac{x}{t_n^{\frac{1}{m}}},\;\; w_0 = u_0 t_n^{\frac{2}{m}},\;\; w_1 = u_1 t_n^{\frac{1}{m}},
\ee
where $m=n+1$ (as shown below).

\smallskip
Defining $u_0(x,t_n)=w_0(z) t_n^{-\frac{2}{m}},\; u_1(x,t_n)= w_1(z) t_n^{-\frac{1}{m}}$, we have
$$
J_0^u =\frac{1}{t_n^{\frac{3}{m}}} J_0^w,\quad J_1^u =  \frac{1}{t_n^{\frac{2}{m}}} J_1^w, \quad  J_2^u = \frac{1}{t_n^{\frac{1}{m}}} J_2^w,
$$
where
\be\label{Jiw}
J_0^w = \pa_z^3+4 w_0 \pa_z+2 w_{0z},\quad  J_1^w = 4 w_1 \pa_z+2 w_{1z},     \quad J_2^w = -4 \pa_z.
\ee
The Hamiltonians have scaling $H_\ell^u(u_0,u_1) = \frac{1}{t_n^{\frac{\ell+1}{m}}} H_\ell^w(w_0,w_1)$.  To fix $m$ we consider the first component of (\ref{dww-triHam}):
\be\label{dww-m=n+1}
u_{0t_n} = J_0^u \delta_{u_1} H_n^u \quad\Rightarrow\quad  -\,\frac{1}{m t_n^{\frac{2}{m}+1}} \, (z w_{0z}+2 w_0) = \frac{1}{t_n^{\frac{n+3}{m}}} \, J_0^w \delta_{w_1} H_n^w \quad\Rightarrow\quad  m = n+1.
\ee
\end{subequations}

Writing (\ref{dww-triHam}) in terms of the scale invariants, we have
\begin{subequations}
\be\label{dww-wt=BdH}
-\frac{1}{n+1} \,\left(\begin{array}{c}
              z w_{0z}+2 w_0  \\
              (z w_1)_z
              \end{array}  \right) = B_2^w \delta_w H_n^w = B_1^w \delta_w H_{n+1}^w = B_0^w \delta_w H_{n+2}^w.
\ee
Looking at the structure of the operators in (\ref{dww-Bi}), we see that the {\em second} row of $B_2^w$ cannot give the expression $(z w_1)_z$, whilst the {\em first} row of $B_0^w$ cannot give the expression $z w_{0z}+2 w_0$, so {\em neither} of these Hamiltonian representations can be reduced to this space.  However, the $B_1^w$ representation {\em is} consistent, giving
\be\label{dww-wt=B1dH}
J_0^w \delta_{w_0} \left(H_{n+1}^w+\frac{z w_0}{2 (n+1)}\right)=0 \quad\mbox{and}\quad \pa_z \delta_{w_1} \left(H_{n+1}^w+\frac{z w_1^2}{8 (n+1)}\right)=0.
\ee
We use the squared eigenfunction (of $J_0^w$) substitution, $\delta_{w_0} \left(H_{n+1}^w+\frac{z w_0}{2 (n+1)}\right) = \frac{1}{2} \varphi^2$ (see (\ref{kdv-ham1w-phi}) and (\ref{kdv-phi-lag})), to obtain the Lagrangian
\be\label{dww-B1-Lag}
{\cal L}_{n+1} = H_{n+1}^w+\frac{z}{8 (n+1)}\, (4 w_0+w_1^2)+\alpha w_1 +\frac{1}{2} (\varphi_z^2 - w_0 \varphi^2) -\frac{\beta}{\varphi^2},
\ee
which is a non-autonomous extension of (4.2) in \cite{f23-1}.
\end{subequations}

\subsection{The Miura map and MDWW Hierarchy}

Whilst there are two modifications of the DWW hierarchy, the operator $B_1^u$ is only associated with the first of these, given by
\begin{subequations}
\be\label{dww-uv-miura}
u_0 = -v_{0x}-v_0^2,\;\; u_1 = v_1, \quad\mbox{with}\quad H_n^v = H_n^u\left( -v_{0x}-v_0^2,v_1\right),
\ee
and
\be\label{dww-mod-Ham}
{\bf u}_{t_n} = B_1^u \delta_u H_{n+1}^u \quad\mbox{gives}\quad {\bf v}_{t_n} = B_1^v \delta_v H_{n+1}^v,\;\;\;\mbox{where}\;\; B_1^v = \left(
                                                                                                                                          \begin{array}{cc}
                                                                                                                                            -\pa_x & 0 \\
                                                                                                                                            0 & 4 \pa_x
                                                                                                                                          \end{array}
                                                                                                                                        \right).
\ee
\end{subequations}
Again, there is an additional integral, $H_{-1}^v = c_0 v_0+c_1 v_1$, which is the Casimir of $B_1^v$.  Also, we can again simplify the form of $H_n^v$ by removing some exact derivatives, which arise in the calculation.

\subsubsection{Reduction through Scaling Symmetry}

The modification has scaling symmetry
\begin{subequations}
\be\label{mdwwscale}
S^v = x \pa_x+ (n+1)t_n \pa_{t_n} - v_0 \pa_{v_0} -v_1 \pa_{v_1},\;\;\mbox{with invariants}\;\; z = \frac{x}{t_n^{\frac{1}{n+1}}},\;\; \theta_0 = v_0 t_n^{\frac{1}{n+1}},\;\; \theta_1 = v_1 t_n^{\frac{1}{n+1}}.
\ee
Writing $v_0(x,t_n) = \theta_0(z) t_n^{-\frac{1}{n+1}},\; v_1(x,t_n) = \theta_1(z) t_n^{-\frac{1}{n+1}}$, equation (\ref{dww-mod-Ham}) for $\bf v$ reduces to
\be\label{dww-mod-Ham-red}
\pa_z \delta_{\theta_0} \left(H_{n+1}^\theta -\frac{z}{2(n+1)} \, \theta_0^2\right) = 0,   \quad
                                     \pa_z \delta_{\theta_1} \left(H_{n+1}^\theta +\frac{z}{8(n+1)} \, \theta_1^2\right) = 0,
\ee
giving the Lagrangian
\be\label{dww-mod-red-lag}
{\cal L}_{n+1}^\theta = H_{n+1}^\theta + \frac{z}{8 (n+1)} \left(\theta_1^2-4 \theta_0^2\right)-\beta_0 \theta_0 - \beta_1 \theta_1.
\ee
\end{subequations}

\subsection{The Case $n=2$}

Here we consider this specific case of the Lagrangians (\ref{dww-B1-Lag}) and (\ref{dww-mod-red-lag}) and the related Hamiltonian systems, together with the Poisson map between them.  The Hamiltonian (\ref{dww-t2hQ}) below is a slight variation on one of the cases in Section 5 of \cite{01-7}.

\subsubsection{The Reduced DWW Case}

Following \cite{f23-1} (see derivation of (4.8a)), we use $-8H_3^w$ in (\ref{dww-B1-Lag}) to obtain
\begin{subequations}
\be\label{dww-t2flow}
{\cal L}_3 = \frac{1}{2} (\varphi_z^2 - w_0 \varphi^2) -\frac{\beta}{\varphi^2} +\alpha w_1 +\frac{z}{24} (4 w_0+w_1^2) + \frac{1}{2}w_{1z}^2-\frac{1}{8} (4 w_0+w_1^2)(4 w_0+5 w_1^2),
\ee
which is degenerate in $w_0$.  Equation $\delta_{w_0}{\cal L}_3 =0$ gives
\be\label{dww-t2flow-w0}
 w_0 = \frac{1}{24} \left(z-3 (\varphi^2+6 w_1^2)\right)  ,
\ee
leading to
\be\label{dww-t2lag}
{\cal L}_3 = \frac{1}{2} \left(w_{1z}^2+\varphi_z^2\right)+\frac{1}{32} \left(16 w_1^4+12 w_1^2 \varphi^2+\varphi^4\right) +\alpha w_1 -\frac{\beta}{\varphi^2}
                          -\frac{z}{48} \left(4 w_1^2+\varphi^2\right) +\frac{z^2}{288}.
\ee
The corresponding Hamiltonian, with $Q_1=w_1,\, Q_2=\varphi,\, P_1 = w_{1z},\, P_2 = \varphi_z$,
\be\label{dww-t2hQ}
h^{(Q)} = \frac{1}{2} (P_1^2+P_2^2)-\alpha Q_1 +\frac{\beta}{Q_2^2} - \frac{1}{32} \left(16 Q_1^4+12 Q_1^2 Q_2^2+Q_2^4\right)  +\frac{z}{48}(4 Q_1^2+Q_2^2)-\frac{z^2}{288} ,
\ee
is a non-autonomous version of (4.10) in \cite{f23-1}.  We can then easily generalise the Poisson commuting function $f^{(Q)}$:
\be\label{dww-t2fQ}
f^{(Q)} = P_2 (Q_2P_1-Q_1P_2) -\frac{1}{2} \alpha\, Q_2^2-2 \beta \,\frac{Q_1}{Q_2^2} -\frac{1}{8} Q_1 Q_2^2 (2 Q_1^2+Q_2^2) +\frac{z}{24}\, Q_1 Q_2^2.
\ee
\end{subequations}

\subsubsection{The Reduced MDWW case}

Using $-8H_3^w$ to calculate $-8H_3^\theta$  in (\ref{dww-uv-miura}), we obtain (removing an exact derivative and multiplying by an overall numerical factor)
\begin{subequations}
\be\label{mdww-t2-lag}
{\cal L}_3 = \frac{1}{8} \left(4 \theta_{0z}^2-\theta_{1z}^2\right) - \frac{3}{4} \theta_1^2 \theta_{0z} + \frac{1}{4} \left(\beta_0 \theta_0+ \beta_1 \theta_1\right)
                     +\frac{1}{32} \left(4 \theta_0^2-\theta_1^2\right) \left(4 \theta_0^2-5\theta_1^2\right) +\frac{z}{96} \left(4 \theta_0^2-\theta_1^2\right).
\ee
Defining $q_1=\theta_0,\, q_2=\theta_1,\, p_1 = \theta_{0z}-\frac{3}{4} \theta_1^2,\, p_2 = -\frac{1}{4} \theta_{1z}$, we obtain the Hamiltonian
\be\label{mdww-t2-hq}
h^{(q)} = \frac{1}{2} p_1^2-2 p_2^2 +\frac{3}{4} q_2^2 p_1 -\frac{1}{8} \left(4 q_1^4-6 q_1^2 q_2^2-q_2^4\right) -\frac{1}{4} (\beta_0 q_1+\beta_1 q_2) - \frac{z}{96} \left(4 q_1^2-q_2^2\right),
\ee
which is a non-autonomous version of (5.8b) in \cite{f23-1}.  We can therefore similarly generalise the Poisson commuting function ($f^{(Q)}$ in section 5.2.1 of \cite{f23-1}):
\bea
f^{(q)} &=& \frac{1}{4} \left(p_1+q_1^2\right) \left(16 q_1p_2+4 q_2 p_1+(8 q_1^2+q_2^2) q_2 \right)+\frac{1}{4} \beta_0 (2 p_2+q_1 q_2) -\frac{1}{2} \beta_1 (p_1+q_1^2) \nn\\
&&      \hspace{1cm}  +\frac{z}{96} \left(4 (q_2p_1+4 q_1 p_2)+q_2 (8 q_1^2+q_2^2)\right).     \label{mdww-t2-fq}
\eea
\end{subequations}

\subsubsection{From Miura Map to Poisson Map}

Given the coordinate definitions, $Q_1 = w_1,\, Q_2 = \varphi,\, P_1 = w_{1z},\, P_2 = \varphi_z$, with constraint $w_0 = \frac{1}{24} \left(z -3 (6 w_1^2+\varphi^2)\right)$, and $q_1=\theta_0,\, q_2 = \theta_1, \, p_1 = \theta_{0z}-\frac{3}{4} \theta_1^2,\, p_2 = - \frac{1}{4} \theta_{1z}$, the Miura map (\ref{dww-uv-miura}) implies
\begin{subequations}
\be\label{dww-mdww-t2-Pmap}
Q_1=q_2,\;\; Q_2 = \sqrt{8 (p_1+q_1^2) +\frac{z}{3}},\;\; P_1 = Q_{1z},\;\; P_2 = Q_{2z}, \quad\mbox{where}\quad Q_{iz} = \{Q_i,h^{(q)}\}.
\ee
To obtain formulae for the {\em dynamical variables} $\alpha$ and $\beta$, we use
\be\label{dww-t2-ab-forms}
\alpha = Q_{1zz} -\frac{1}{4} Q_1 \left(8 Q_1^2+3 Q_2^2\right)+\frac{z}{6} \, Q_1,\quad \beta = \frac{Q_2^3}{48} \left(24 Q_{2zz} - 3 Q_2 \left(6 Q_1^2+Q_2^2\right)+ z Q_2\right),
\ee
where $Q_{izz} = \{\{Q_i,h^{(q)}\},h^{(q)}\}$.

\medskip
Using (\ref{mdww-t2-hq}), we obtain  {\small
\be\label{dww-t2-Pmap}
Q_1=q_2,\;\; Q_2 = \sqrt{8 (p_1+q_1^2) +\frac{z}{3}},\;\; P_1 = -4p_2,\;\; P_2 = \frac{\beta_0+q_1 \left(8(p_1+q_1^2)+\frac{z}{3}\right)}{\sqrt{8(p_1+q_1^2)+\frac{z}{3}}},
                  \;\;\alpha = -\beta_1 ,\quad \beta = -\frac{1}{2} \beta_0^2 .
\ee
}  With this transformation, we have
\be\label{hQhq}
h^{(Q)} = -4 h^{(q)},\quad f^{(Q)} = -8 f^{(q)}+\frac{1}{6} \beta_1 z.
\ee
\end{subequations}

\subsubsection{The Poisson Brackets}

As with the original DWW hierarchy, this system is tri-Hamiltonian.  In the $6-$dimensional space with coordinates ${\bf Q} = (Q_1,Q_2,P_1,P_2,\beta, \alpha)$, the Poisson matrices are given by
\begin{subequations}
{\small
\bea
&&  {\cal P}_0^{(Q)} =  \left(
                     \begin{array}{cccccc}
                       0 & 0 & 0 & \frac{1}{Q_2} & a_{15} & 0 \\[2mm]
                       0 & 0 & \frac{1}{Q_2} & -\frac{2 Q_1}{Q_2^2} & a_{25} & 0 \\[2mm]
                       0 & -\frac{1}{Q_2} & 0 & \frac{P_2}{Q_2^2} & a_{35} & 0 \\[2mm]
                        -\frac{1}{Q_2} & \frac{2 Q_1}{Q_2^2} & - \frac{P_2}{Q_2^2} & 0 & a_{45} & 0 \\[2mm]
                      -a_{15} & -a_{25} & -a_{35} & -a_{45} & 0 & 0\\[2mm]
                       0 & 0 & 0 & 0  & 0 & 0
                     \end{array}
                   \right), \quad
{\cal P}_1^{(Q)} = \left(
                     \begin{array}{rrrrrr}
                       0 & 0 & 1 & 0 & 0 & 0  \\
                       0 & 0 & 0 & 1 & 0 & 0 \\
                       -1 & 0 & 0 & 0 & 0 & 0 \\
                        0 & -1 & 0 & 0 & 0 & 0 \\
                       0 & 0 & 0 & 0 & 0 & 0\\
                       0 & 0 & 0 & 0 & 0 & 0
                     \end{array}
                   \right), \nn\\
                   &&     \label{dww-t2-PBs}     \\
&&  {\cal{P}}^{(Q)}_2 =
                  \left(\begin{array}{cccccc}
                 0&0&2 Q_1& Q_2 & 0 & b_{16}\\[2mm]
  0&0& Q_2&0&0 &b_{26}\\[2mm]
  -2Q_1&-Q_2&0&P_2&0 &b_{36}\\[2mm]
  -Q_2&0&-P_2&0&0 &b_{46}\\[2mm]
  0&0&0&0&0&0\\[2mm]
  -b_{16}&-b_{26}&-b_{36}&-b_{46}&0&0
                              \end{array}\right) ,    \nn
\eea
}where \; $(a_{15},a_{25},a_{35},a_{45},0,0)^T= -{\cal{P}}_1^{(Q)} \nabla_{Q}f^{(Q)}$ \; and  \;
$(b_{16},b_{26},b_{36},b_{46},0,0)^T=2{P}_1^{(Q)}\nabla_{Q}h^{(Q)}$.  In fact these formulae are just those of (5.10d) in \cite{f23-1}.  Some non-autonomous parts are hidden within the definition of $h^{(Q)}$ and $f^{(Q)}$, used to define $a_{i5}$ and $b_{i6}$.

\smallskip
Each of these Poisson brackets has two Casimirs and can be used to generate two commuting flows:
\bea
&&  {\cal P}_0^{(Q)}\nabla_Q h^{(Q)} = {\cal P}_0^{(Q)}\nabla_Q\, \alpha = {\cal P}_1^{(Q)}\nabla_Q \,\beta = {\cal P}_1^{(Q)}\nabla_Q \,\alpha = {\cal P}_2^{(Q)}\nabla_Q f^{(Q)} = {\cal P}_2^{(Q)}\nabla_Q \,\beta = 0, \nn\\
&&  {\bf Q}_{t_h} = {\cal P}_2^{(Q)}\nabla_Q \left(\frac{1}{2} \alpha\right) = {\cal P}_1^{(Q)}\nabla_Q h^{(Q)} = {\cal P}_0^{(Q)}\nabla_Q f^{(Q)},  \label{dww-t2-Qthtf}    \\
&&  {\bf Q}_{t_f} = {\cal P}_2^{(Q)}\nabla_Q h^{(Q)} = {\cal P}_1^{(Q)}\nabla_Q f^{(Q)} = {\cal P}_0^{(Q)}\nabla_Q (-\beta).  \nn
\eea
\end{subequations}

\medskip
Using the map (\ref{dww-t2-Pmap}), we can transform the Poisson brackets (\ref{dww-t2-PBs}) to the $\bf q$ space, with
$$
\frac{\pa {\bf q}}{\pa {\bf Q}}\, {\cal P}_0^{(Q)} \left(\frac{\pa {\bf q}}{\pa {\bf Q}}\right)^T = -\frac{1}{4}{\cal P}_0^{(q)}, \quad
          \frac{\pa {\bf q}}{\pa {\bf Q}}\, {\cal P}_1^{(Q)} \left(\frac{\pa {\bf q}}{\pa {\bf Q}}\right)^T = -\frac{1}{4}{\cal P}_1^{(q)},\quad
                    \frac{\pa {\bf q}}{\pa {\bf Q}}\, {\cal P}_2^{(Q)} \left(\frac{\pa {\bf q}}{\pa {\bf Q}}\right)^T = \frac{1}{2}{\cal P}_2^{(q)},
$$
where
\begin{subequations}     {\small
\bea
&&  {\cal P}_0^{(q)} =   \frac{1}{4\beta_0} \left(\begin{array}{cccccc}
                           0& -16q_1 & 8(2 p_2+q_1q_2) & 4p_1+8 q_1^2+3 q_2^2 & a_{15} & 0\\
                           16q_1 & 0 & -32 q_1^2 & 0 & a_{25} & 0\\
                             -8(2 p_2+q_1q_2) & 32 q_1^2 & 0 & a_{34} & a_{35} & 0\\
                             -(4p_1+8 q_1^2+3 q_2^2) & 0 & -a_{34} & 0 & a_{45} & 0  \\
                             -a_{15} & - a_{25} & - a_{35} & -a_{45} & 0 & 0  \\
                              0 & 0 & 0 & 0 & 0 & 0
                              \end{array}\right), \nn\\
                   &&     \label{mdww-t2-PBs}     \\
&& {\cal P}_1^{(q)} = \left(
                     \begin{array}{rrrrrr}
                       0 & 0 & 1 & 0 & 0 & 0  \\
                       0 & 0 & 0 & 1 & 0 & 0 \\
                       -1 & 0 & 0 & 0 & 0 & 0 \\
                        0 & -1 & 0 & 0 & 0 & 0 \\
                       0 & 0 & 0 & 0 & 0 & 0\\
                       0 & 0 & 0 & 0 & 0 & 0
                     \end{array}
                   \right),  \;\;
 {\cal{P}}^{(q)}_2 =
                  \left(\begin{array}{cccccc}
                 0 & -2 & 0 &  q_1 & 0 & b_{16}\\[1mm]
                  2 & 0 & -4 q_1 & - q_2 & 0 & b_{26} \\
                  0 & 4 q_1 & 0 & b_{34} & 0 & b_{36}  \\
                  - q_1 &  q_2 & -b_{34} & 0 & 0 & b_{46}  \\
                  0 & 0 & 0 & 0 & 0 & 0\\
                  -b_{16} & -b_{26} & -b_{36} & -b_{46} & 0 & 0
                       \end{array}\right),     \nn
\eea
}where \;$a_{34} = \beta_0-2 q_1 (4p_1+8 q_1^2+3 q_2^2)$,\; $(a_{15},a_{25},a_{35},a_{45},0,0)^T= -32{\cal{P}}_1^{(q)} \nabla_{q}f^{(q)}$,\; $b_{34}=  -\left(p_1+3 q_1^2+\frac{z}{24}\right)$, \; and  \;
$(b_{16},b_{26},b_{36},b_{46},0,0)^T=-4{\cal{P}}_1^{(q)}\nabla_{q}h^{(q)}$.  These are the same formulae as the Poisson brackets of the first modification of the DWW case, given in Section 5.2.2 of \cite{f23-1}.  Again, some non-autonomous parts are hidden within the definition of $h^{(q)}$ and $f^{(q)}$, used to define $a_{i5}$ and $b_{i6}$.

\smallskip
Each of these Poisson brackets has two Casimirs and can be used to generate two commuting flows:
\bea
&&  {\cal P}_0^{(q)}\nabla_q h^{(q)} = {\cal P}_0^{(q)}\nabla_q\, \beta_1 = {\cal P}_1^{(q)}\nabla_q \,\beta_0 = {\cal P}_1^{(q)}\nabla_q \,\beta_1
               = {\cal P}_2^{(q)}\nabla_q \,\beta_0= {\cal P}_2^{(q)}\nabla_q \left(f^{(q)} - \frac{\beta_1}{48}\, z\right) = 0, \nn\\
&&  {\bf q}_{t_h} = {\cal P}_2^{(q)}\nabla_q \left(-\frac{1}{4} \beta_1\right) = {\cal P}_1^{(q)}\nabla_q h^{(q)} = {\cal P}_0^{(q)}\nabla_q (2f^{(q)}),  \label{dww-t2-qthtf}    \\
&&  {\bf q}_{t_f} = {\cal P}_2^{(q)}\nabla_q (-h^{(q)}) = {\cal P}_1^{(q)}\nabla_q f^{(q)} = {\cal P}_0^{(q)}\nabla_q \left(-\frac{1}{16}\beta_0^2\right).  \nn
\eea
\end{subequations}

\section{Conclusions}

The purpose of this paper has been to show that the Painlev\'e reductions of the cKdV hierarchy fit neatly into the framework developed in \cite{f23-1,f24-1}, with very little additional calculation.

In the two examples explained in this paper, just one of the Hamiltonian operators (of the PDE hierarchy) could be reduced to give a Lagrangian formulation for the Painlev\'e reduction.  However, {\em all} the Poisson brackets of the stationary reduction are directly applicable the Painlev\'e case, {\em unchanged in structure}.  In detail, there are differences, but these are hidden in the form of the Poisson commuting functions.

These calculations could easily be generalised to the cases of higher $N$ (in (\ref{Scrod-op})), as in \cite{f24-1}.


\end{document}